%% file: CSCISecondDraftV3.tex
\documentclass[10pt, conference, compsocconf]{IEEEtran}
\IEEEoverridecommandlockouts

\usepackage[toc,acronym]{glossaries}
\usepackage{adjustbox}
\usepackage{tikz}
\usetikzlibrary{arrows,shadows,backgrounds,calc,decorations.pathreplacing,decorations.pathmorphing,positioning, 
	automata}
\usepackage{pgfplots}
\pgfplotsset{compat=1.13} 
\usepackage{pgfplotstable}
\makeglossaries
\usepackage{listings}

\newacronym{ANN}{ANN}{Artificial Neural Network}
\newacronym{FPGA}{FPGA}{Field Programmable Gate Array}  
\newacronym{HW}{HW}{hardware}
\newacronym{ISA}{ISA}{Instruction Set Architecture}
\newacronym{I/O}{I/O}{Input/Output}
\newacronym{MC}{MC}{Multi-Core and/or Many-Core}
\newacronym{MLP}{MLP}{Memory Level Parallelism}
\newacronym{OoO}{OoO}{Out-of-Order}
\newacronym{OS}{OS}{operating system}
\newacronym{PD}{PD}{Propagation Delay}
\newacronym{PU}{PU}{Processing Unit}
\newacronym{SPA}{SPA}{Single Processor Approach}
\newacronym{SW}{SW}{software}
\newacronym{HPL}{HPL}{High Performance Linpack}
\newacronym{HPCG}{HPCG}{High Performance Conjugate Gradients}

\definecolor{webgreen}{rgb}{0,.5,0}
\definecolor{webbrown}{rgb}{.6,0,0}
\definecolor{webyellow}{rgb}{0.98,0.92,0.73}
\definecolor{webgray}{rgb}{.753,.753,.753}
\definecolor{webblue}{rgb}{0,0,.8}
\definecolor{webgreen}{rgb}{0, 0.5, 0} 
\definecolor{webred}{rgb}{0.5, 0, 0}   

\usepackage{cite}
\usepackage{amsmath,amssymb,amsfonts}
\usepackage{graphicx}
\usepackage{textcomp}
\usepackage{xcolor}

\begin{document}
	
	\title{von Neumann's missing "Second Draft":\\
		what it should contain\\
		\thanks{Project no. 136496  has been implemented with the support provided from the National Research, Development and Innovation Fund of Hungary, financed under the K funding scheme.
			Submitted to 2020 International Conference on Computational Science and
			Computational Intelligence, Las Vegas, US, as paper CSCI2019
		}
	}
	
	\author{\IEEEauthorblockN{J\'anos  V\'egh}
		\IEEEauthorblockA{\textit{Kalim\'anos BT} \\
			Debrecen, Hungary \\
			Vegh.Janos@gmail.com~ORCID:~0000-0002-3247-7810}
	}

	\maketitle
	
	\begin{abstract}
		Computing science is based on a computing paradigm that is not valid anymore for today's technological conditions. The reason is that the transmission time even inside the processor chip, but especially between the system's components, is not negligible anymore. The paper introduces a quantitative measure for dispersion, which is vital for computing performance and energy consumption, and demonstrates how its value increased with the changing technology. The temporal behavior (including the dispersion of the commonly used synchronization clock time) of computing components has a critical impact on the system's performance at all levels, as demonstrated from gate-level operation to supercomputing. The same effect limits the utility of the researched new materials/effects if the related transfer time cannot be proportionally mitigated.
		Von Neumann's model is perfect, but now it is used outside of its range of validity.
		The correct procedure to consider the transfer time for the present technological background is also derived.
	\end{abstract}
	

	\section{Introduction}\label{sec:Introduction}
	
	Today's computing is commonly thought to be based on the famous "First Draft"~\cite{EDVACreport1945} by von Neumann\footnote{This publication is commonly referred to as "von Neumann architecture", although von Neumann in the first sentence makes clear that "\textit{The considerations which follow deal with the \textbf{structure} of a very high speed automatic digital computing system, and in particular with its \textbf{logical control}.}" }.
	In that report, he discussed the principles and the technical implementation of his paradigm in parallel with the brain's neuronal operation,
	so his model was brain-inspired.
	He emphasized the computing process's timing relations and analyzed the role of delays in the computing chain.
	He also called attention to the fact that \textit{the transfer time is an integral part of the computation process}: whether it is biological, digital, or analog,
	processing cannot even begin before its input operands are delivered
	to the place of operation, and vice versa, the output operand cannot be delivered until the computing process terminates.
	\textit{The transfer time and the operation time mutually block each other in a computing chain}.
	
	Given that \textit{he considered the intended vacuum tube implementation}, to simplify the model, \textit{he proposed to neglect the
		transfer time}, aside from the processing time.
	In other words, his approximation assumes instant interaction between his "computing organs".
	However, he explicitly warned that it would be \textit{unsound} to use that classic paradigm, neglecting the conduction time,
	to describe neuronal operation, given that the conduction time is longer than the synaptic time.
	
	Another condition that --with his words-- \textit{vitiates} his paradigm, is if the processor is "too fast".
	If we consider a 300~$m^2$ sized computer room and the 3000 vacuum tubes estimated, von Neumann considered a distance between vacuum tubes about 30~cm as a critical value. At this distance, the transfer time is about three orders of magnitude lower than the processing time (between the "steps" he mentioned). These limitations are why von Neumann justified the procedure \textit{for vacuum tube technology only}. He noted that
	using a hundred-fold higher frequency, even with vacuum tubes, \textit{vitiates the neglection} he proposed. 
	At such a frequency, the transfer time approaches (or even exceeds) the order of magnitude of the processing time, so neglecting the transfer time cannot anymore be justified: the \textit{apparent processing time}  (the clock time between consecutive computing operations) differs from the 
	\textit{physical processing time}  by dozens of percent.
	
	\section{The history of temporal characteristics}
	
	Today, the size of a processor chip is in the three~cm range, and correspondingly, the transfer time can reach the 1~ns range. The processing time (in the sense as von Neumann used the term) is well below the range of 1~ns,
	so it is worth to check if today's computing can be based on the classic paradigm.
	"\textit{Building this new hardware  [neuromorphic computing] necessitates reinventing electronics}";
	"\textit{more physics and materials needed}"~\cite{PhysicsForNeuromorhicComputing:2020}.
	It is not sufficient, however. At least \textit{\textbf{some physics is needed even in the computing paradigm}}, to benefit from the researched new effects and materials.
	It was noticed that "\textit{computers have undergone tremendous improvements in performance over the last 60 years, but those improvements have significantly
		slowed down over the last decade, owing to fundamental limits in the underlying computing primitives}"~\cite{BrainInspiredBlocks:2020}.
	However, the real reason is at one level deeper: \textit{we are using the otherwise correct computing model under technological conditions where the neglections used in the classic paradigm are not valid any more}.

	\begin{figure}[t!]
		\includegraphics[width=1.\columnwidth]
		{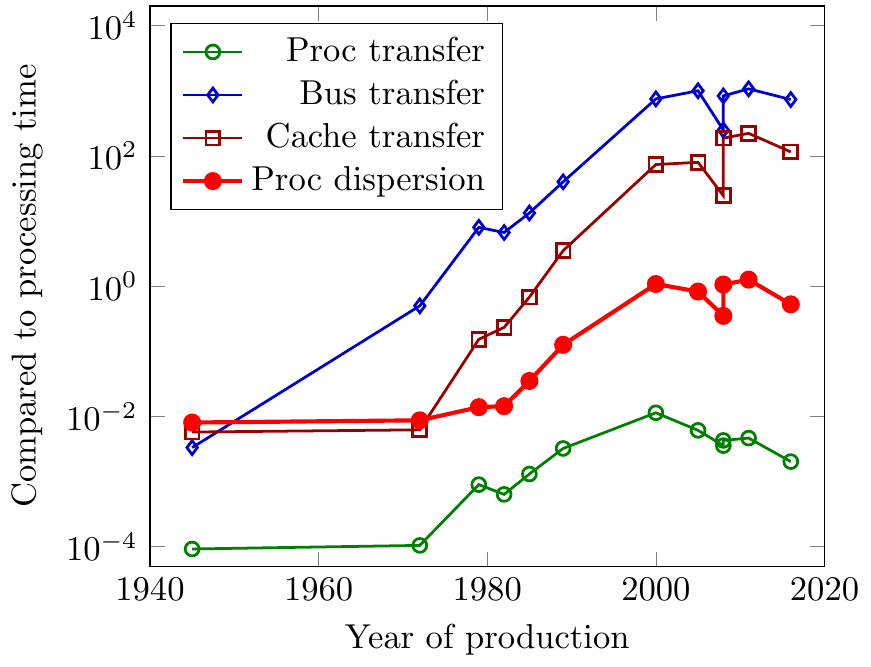}\vspace{-\baselineskip}
		\label{ProcessorDispersion}
		\caption{The history of some relative temporal characteristics of processors, in function of their year of production. Notice how cramming more transistors in a processor changed disadvantageusly their temporal characterisctics.\label{ProcessorDispersion}
			\vspace{-\baselineskip}}
	\end{figure}
	
	Given that von Neumann derived his neglections \textit{apart from processing time}, we derive some quantitative merits, as discussed below. Fig.~\ref{ProcessorDispersion} shows their dependence on the year of fabrication of the processor. The technical data are taken from publicly available data\footnote{$https://en.wikipedia.org/wiki/Transistor\_count$} and from~\cite{EDVACEckertMauchly}.
	The figures of the merits are results of rough and somewhat arbitrary approximations. However, their consequent use enables us to draw limited validity conclusions without needing proprietary technological data.
	
	We estimate the distance between the processing elements in two different ways. We use the method described above for vacuum tubes to calculate the square root of the processor area divided by the number of transistors. This figure gives a kind of "average distance" of the transistors. We consider it as a minimum distance the signals must travel between transistors\footnote{Notice that this transfer time also shows a drastic increase with the number of transistors, but alone does not vitiate the classic paradigm}. This value is depicted as "Proc transfer" in Fig.~\ref{ProcessorDispersion}.   The maximum is the distance between the two farthest processing elements on the chip\footnote{Evidently, introducing clock domains and multi-core processors, shades the picture. However, we cannot provide a more accurate estimation without proprietary technological data}.

	Given that usually the processing elements and the storage elements are fabricated as separated technological blocks and connected by wires (aka bus), we also estimated a "bus transfer" time.
	The memory access in this way is extended by the bus transfer time. We assumed that a cache memory could be advantageously positioned at a(n average) distance of half processor size because of this effect. This time is shown as "Cache transfer" time. The cache memories appeared about the end of the 1980s. It became evident that the bus transfer vastly increases the memory transfer time. Using cache memory can enhance systems' performance by order of magnitude (cache data can be calculated for all processors, however).
	
	Given that "The emphasis is on the exclusion of a dispersion"~\cite{EDVACreport1945}, 
	using those technical data, we define a dispersion as the
	geometric mean of the minimum and maximum "Proc transfer" times, divided by the processing time.
	In von Neumann's abstraction, the "well-defined dispersionless synaptic delay $\tau$ [processing time]" is used. With our definition, the dispersion of EDVAC is (at or below) 1~\%; this is why von Neuman justified his procedure. 
	An interesting parallel is that both EDVAC and Intel 8008 have the same number of processing elements. The relative processor and cache transfer times are in the same order of magnitude.
	However, notice that the bus transfer time's importance has grown and started to dominate the single-processor performance in personal computers.
	A decade later, the physical size of the bus necessitated to introduce cache memories. 
	The physical size led to saturation in all relative transfer times.\footnote{The real cause of the "end of the Moore age", is, that Moore's observation is not valid for the physical bus size} The slight decrease in the relative times in the past years can probably be attributed
	to the sensitivity of our calculation method to the spread of multi-cores; this suggests to repeat our analysis method with proprietary technological data.
	
	As the "Proc dispersion" diagram line shows, \textit{in the today's technology, the dispersion is near to unity}. This large dispersion means that today's processors' operating regime is more close to the operating regime
	of our brain than the operating model abstracted in the classic paradigm.  However, for our brain, an explicit "spatiotemporal" behavior is considered, and is "unsound" to use the classic paradigm to describe it. That is, we cannot apply the "dispersionless" classic paradigm any more\footnote{Reaching the plateau of the diagram lines coincides with introducing the "explicitly parallel instruction set computer"~\cite{EPIC:2000}:
		that was the maximum that the classic paradigm enabled.}. 
	This high dispersion value is why only a small fragment of the input power can be used for computation; the rest is dissipated (produces heat). We experience it since the dispersion approached unity about two decades ago.
	\textit{The dispersion of synchronizing the computing operations} vastly increases the cycle time,  decreases the utilization of all computing units, and enormously increases the power consumption of computing~\cite{ClockDistribution:2012,DarkSilicon2012}.

	\begin{figure*}[t!]
		\vspace{-1.5cm}
		\includegraphics[width=.95\textwidth]
		{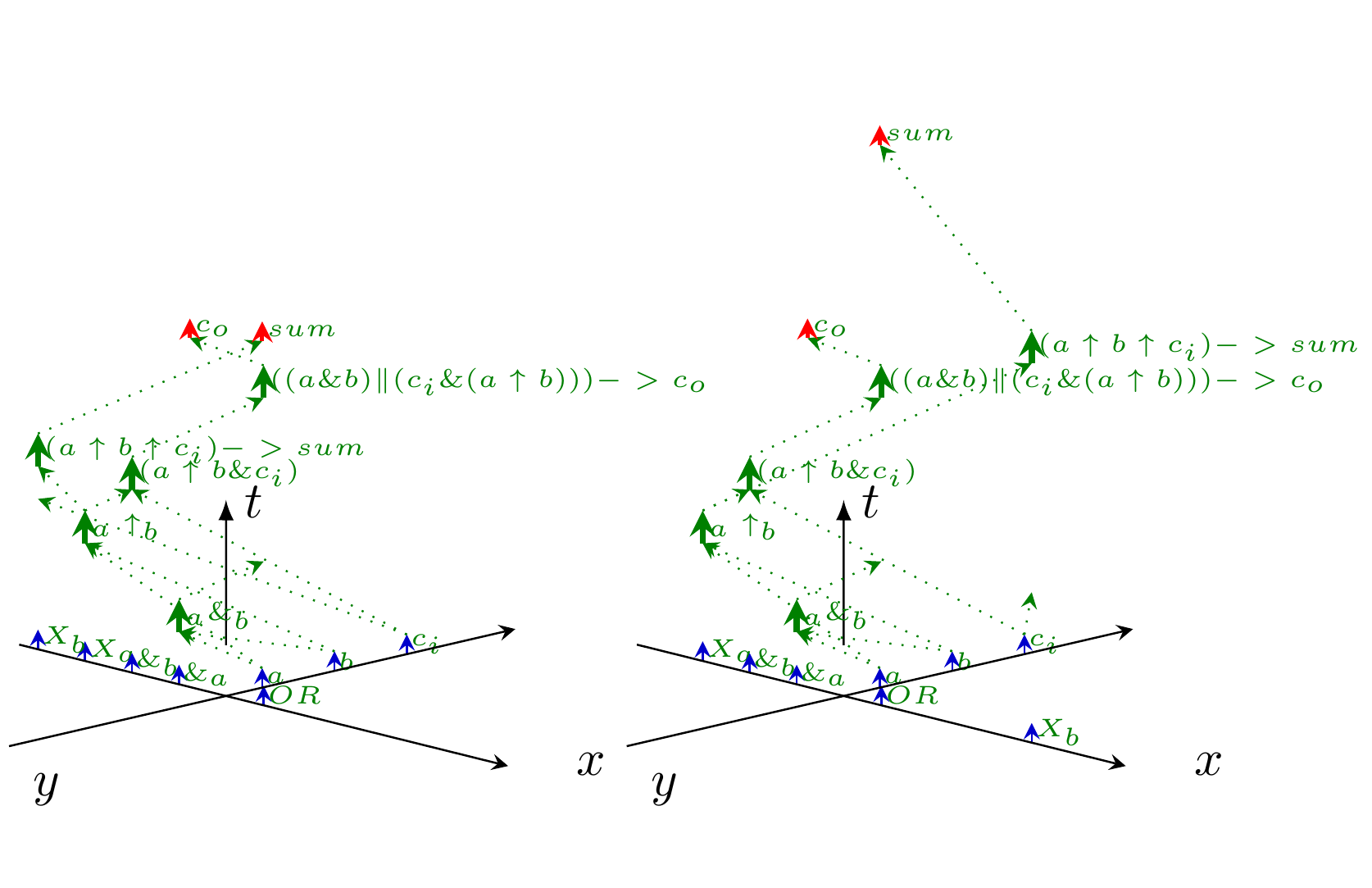}\vspace{-2\baselineskip}
		\label{OneBitAdder}
		\caption{The temporal diagram of a one-bit adder in the time-space system. The diagram shows the logical equivalent of the SystemC source code of Listing~\ref{lst:OneBitAdder},
			\textit{the time from axis $x$ to the bottom of green arrows} signals "idle waiting" time (undefined gate output). In the left subfigure, the second XOR gate is at (-1,0). In the right subfigure, the second XOR gate is at (+1,0). Notice how changing the position of a gate affects signal timing. Notice also that the lack of vertical green arrows means idle time for the gates.\label{OneBitAdder}}
	\end{figure*}

	In the same section, von Neumann said: "We propose to use the delays $\tau$ as absolute units of time which
	can be relied upon to synchronize the functions of various parts of the device. The advantages of
	such an arrangement are immediately plausible".
	Yes, his statement is valid for the well-defined \textit{dispersionless synaptic delay $\tau$} he assumed, but not at all for today's processors. The recent activity~\cite{PhysicsForNeuromorhicComputing:2020,BrainInspiredBlocks:2020} to consider asynchronous operating modes
	is motivated by admitting that \textit{the present synchronized working method is disadvantageous in the non-dispersionless world}.
	\index{asynchronous operating mode}
	
	At his time and in the age of vacuum tube technology, von Neumann did not feel the need to discuss what a procedure can justify describing the computing operation in a non-dispersionless case. However, he suggested reconsidering the validity of the neglections he used in his paradigm for any new future technology. To have a firm computing paradigm for the present technologies, \textit{we need to consider the ratio of the transfer time to processing time}; we cannot neglect it anymore. The real question is, the discussion of which is \textit{missing from the "First draft", what a procedure shall be followed if the transfer time is not negligible}?

	\section{Introducing temporal logic into computing}
	Although von Neumann explicitly mentioned that 
	the propagation speed of electromagnetic waves 
	limits the operating speed of the electronic components,
	until recently, that effect was not admitted in computing. 
	In contrast, in biology, the "spatio-temporal" behavior
	was recognized very early.
	The recent trend is to describe theoretically and model the neuronal operation electronically using the computing paradigm, proposed by von Neumann, which is undoubtedly not valid for today's technology.
	According to von Neumann, it is doubly $unsound$ if one attempts to mimic neural operation based on a paradigm that is $unsound$ for that goal, on a technological base (other than vacuum tubes) that $vitiates$ the paradigm.
	
	Fortunately, the spatio-temporal behavior suggests a "procedure" that can be followed in the case when the transfer time can even be longer than the processing time. Despite the name "spatio-temporal", biology describes its systems' behavior using separated space and time functions (and, as a consequence, needs ad-hoc suggestions and solutions for different problems). However, it has one common attribute with technical computing: the information transfer speed is limited in both of them. For the first look, it seems to be strange to describe such systems with (the inverse of) the Minkowski transform, given that it became famous in connection with Einstein's theory of special relativity. However, in its original form, only the existence of a limiting speed is assumed.
	
	As discussed in~\textbf{\cite{BiologySpatioTemporal:2020}},
	this latter feature enables us to describe
	the correct behavior of information processing both in science-based technical implementations and biology,
	for any combination of the transfer time and the processing time. The key idea is to transform 
	the spatial distances between computing components (that can be $Si$ gates, cores, network nodes, biological or artificial neurons) to time (measured with the limiting speed along the signal path). 
	
	\begin{figure*}
		\includegraphics[width=.95\textwidth]
		{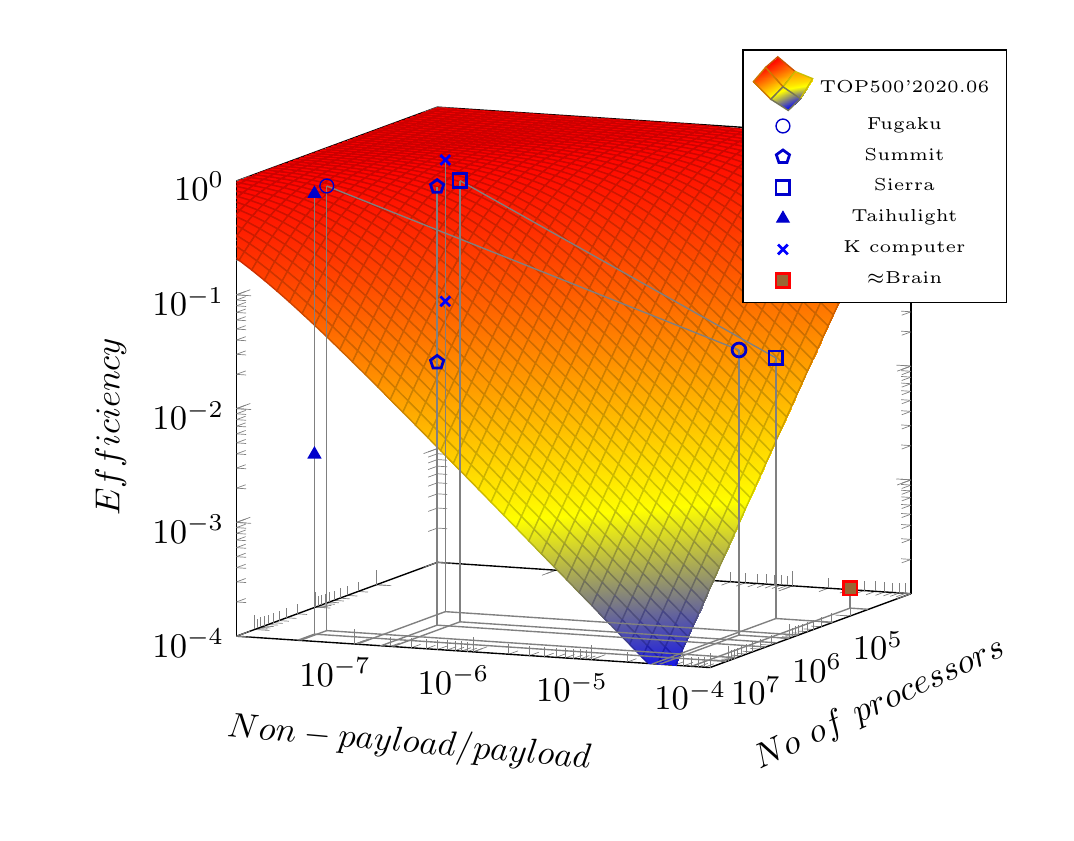}
		\vspace{-\baselineskip}
		\caption{The surface and the figure marks show at what efficiency the top supercomputers run the
			'best workload' benchmark \gls{HPL}, and the 'real-life load' \gls{HPCG}.~\textbf{\cite{VeghHowMany:2020}}
			The right bottom part displays the expected efficiency~\textbf{\cite{VeghBrainAmdahl:2019}} of running
			neuromorphic calculations on \gls{SPA} computers.\label{fig:EfficiencyIdleTime}}
		\vspace{-\baselineskip}
	\end{figure*}

	\section{Phenomena due to the temporal behavior}
	
	Yes, in computing, "more physics \dots is needed"~\cite{PhysicsForNeuromorhicComputing:2020}.	
	
	\subsection{One-bit adder}
	
	\begin{lstlisting}[float,caption=The essential lines of source code of the one-bit adder implemented in 
	SystemC,label=lst:OneBitAdder]
	//We are making a 1-bit addition
	aANDb = a.read() & b.read();
	aXORb = a.read() ^ b.read();
	cinANDaXORb = cin.read() & aXORb;
	
	//Calculate sum and carry out
	sum = aXORb ^ cin.read();
	cout = aANDb | cinANDaXORb;
	\end{lstlisting}
	
	Describing the temporal operation at gate level is an excellent example, that \textit{the line-by-line compiling
		(sequential programming, called also Neumann-style programming~\cite{BackusNeumannProgrammingStyle}),
		formally introduces only logical dependence, but through its technical implementation
		it implicitly and inherently introduces a temporal behavior, too}.
	Fig.~3 in~\cite{EDVACreport1945} shows a simple adder,
	in vacuum-tube approach. 
	Listing~\ref{lst:OneBitAdder} shows how a one-bit adder is implemented 
	in \gls{HW}-description language SystemC.
	Fig.~\ref{OneBitAdder} shows the corresponding elementary operations,
	as they happen along the axis $t$ (for a detailed explanation and legend, see~\textbf{\cite{VeghTemporal:2020}}). Notice that considering their temporal
	behavior results in longer delays for the adder built 
	from gates used in today's technology, than that for the logical adder
	assumed by von Neumann.
	Notice also, what von Neumann called the attention to: \textit{until a gate receives
		all of its inputs, its output is undefined}. For the drastic consequences of this effect, see the impact of training \gls{ANN}s in~\textbf{\cite{VeghScalingANN:2020}}.
	
	\begin{figure*}
		\begin{tabular}{cc}
			\includegraphics[width=1.1\columnwidth]
			{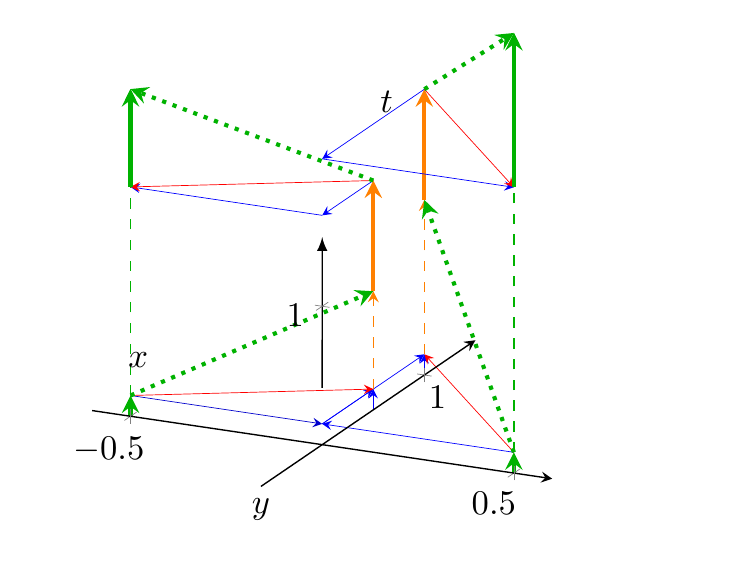}
			&\hspace{-2cm}
			\includegraphics[width=1.2\columnwidth]
			{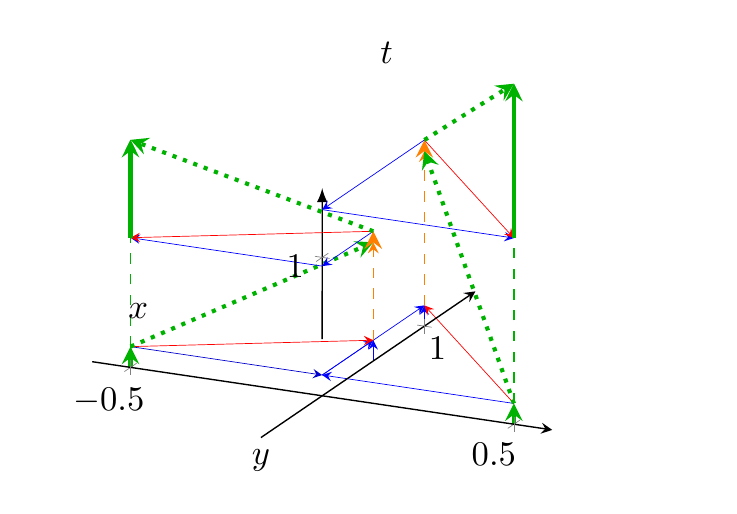}
		\end{tabular}	
		\caption{The performance dependence of an on-chip cache memory, at different cache operating times, in the same topology.
			The cores at x=-0.5 and x=0.5 positions access the on-chip cache at y=0.5 and y=1.0, respectively. The vertical orange arrows represent the physical cache operating time, and vertical the green arrows the apparent access time. The physical operating speed of the cache memory of the right subfigure is ten times better. Compare the apparent access times to the corresponding physical ones (the time ratio is better only about a factor of two). Notice also that the apparent operating speed is more sensitive to the position rather than to the speed of the cache memory\label{fig:CachePerformance}}
		\vspace{-\baselineskip}	
	\end{figure*}
	
	\subsection{The inherent performance limits of parallelized computing}
	The temporal behavior of the components leads to a drastic performance loss
	in the case of parallelized computing.
	Recall that "\textit{this decay in performance is not a fault of the architecture, but is dictated by the limited parallelism}".\cite{ScalingParallel:1993}
	Fig.~\ref{fig:EfficiencyIdleTime} shows the efficiency of some recent supercomputers in function of the number of their processor cores,
	and their degree of parallelization~\textbf{\cite{VeghHowMany:2020}}.  Notice how the different workload (\gls{HPCG}) forces to use only a fragment of the available cores.
	
	\subsection{New effects and materials}
	Their temporal behavior also limits the usability of new materials/effects, a highly popular idea, especially for neuronal operations~\cite{PhysicsForNeuromorhicComputing:2020,NatureBuildingBrain:2020}. 
	Fig.~\ref{fig:CachePerformance} shows the effect of using a faster cache memory in a computing system. Although the apparent operation gets faster
	when using a ten-fold quicker (and maybe even more expensive) cache, 
	the reached speedup is not proportional to the speedup in the cache's physical operating speed.
	For a detailed discussion see~\textbf{\cite{VeghTemporal:2020}}.
	
	\section{Summary}
	The present commonly used classic computing paradigm, according to its inventor, is valid for (the timing relations of)
	vacuum tubes only. It assumes instant interaction between
	the components of computing systems, which is theoretically wrong and contradicts everyday experience. Concerning their temporal behavior, the computing systems show interesting (but not surprising) parallels with the modern science~\textbf{\cite{VeghModernParadigm:2019,VeghBrainAmdahl:2019}}.
	The phenomena range from the stalled
	single-processor (used for heating rather than computing) performance through the supercomputers, unable to exceed their inherent performance rooflines~\textbf{\cite{VeghHowMany:2020}} and having \textit{payload performance} for real-life task about 1~\% of their \textit{nominal performance}, to
	the stalled scaling of \gls{ANN}s~\textbf{\cite{VeghScalingANN:2020}} (the latter led to that
	"AI Core progress has stalled"~\cite{AIcoreProgressStalled:2020}). Yes, 
	"\textit{building this new hardware necessitates reinventing electronics}"~\cite{PhysicsForNeuromorhicComputing:2020}.
	
	At this point, we have two options. We can complement the currently existing names for the scientific discipline and its journals so that "Computing Science \textit{for vacuum tubes only}". Furthermore, we can start a new discipline under the name "Modern computing science" (parallel with classic versus modern science) based on temporal logic, describing the computing based on state-of-the-art technology.
	In this way, we can enter into the "Next Level"~\cite{ComputingPerformance:2011}.
	Even \textit{rebooting computing} is not possible without
	making such a drastic change.
	The other option is to keep the classic computing science
	and to admit that the "Game Is Over".
	\input{CSCISecondDraftV3.bbl}

	
\end{document}